\documentclass[traditabstract]{aa}
\usepackage{txfonts}
\usepackage{epsfig}
\usepackage{lscape}
\bibpunct{(}{)}{;}{a}{}{,}    %% natbib format like A&A and ApJ
\usepackage{graphicx,graphics}
\usepackage{color}
\usepackage{ulem}
\usepackage{amsmath}
\usepackage{amssymb}
\usepackage[breaklinks=true, colorlinks=true, urlcolor=blue, citecolor=blue, linkcolor=blue]{hyperref}

\begin{document}

\title{The dependence of the gradients of oxygen and nitrogen-to-oxygen on stellar age in MaNGA galaxies}

\author{
        I.~A.~Zinchenko\inst{\ref{LMU},\ref{MAO}} \and
        J.~M. V\'{i}lchez\inst{\ref{IAA}} \and
        E.~P\'{e}rez-Montero\inst{\ref{IAA}} \and 
        A.V.~Sukhorukov\inst{\ref{MAO},\ref{IAC},\ref{ULL}} \and
        M.~Sobolenko\inst{\ref{MAO}} \and
        S.~Duarte Puertas\inst{\ref{LU},\ref{IAA}} 
       }
       
\institute{
Faculty of Physics, Ludwig-Maximilians-Universit\"{a}t, Scheinerstr. 1, 81679 Munich, Germany \label{LMU}
\and
Main Astronomical Observatory, National Academy of Sciences of Ukraine, 
27 Akademika Zabolotnoho St., 03143, Kyiv, Ukraine\label{MAO}
\and
Instituto de Astrof\'{i}sica de Andaluc\'{i}a (CSIC), Apartado 3004, 18080 Granada, Spain \label{IAA} 
\and
D\'epartement de Physique, de G\'enie Physique et d’Optique, Universit\'e Laval, and Centre de Recherche en Astrophysique du Qu\'ebec (CRAQ), Qu\'ebec, QC, G1V 0A6, Canada
\label{LU}
\and
Instituto de Astrof\'{\i}sica de Canarias, E-38205 La Laguna, Tenerife, Spain
\label{IAC}
\and
Dpto.\ de Astrof\'{\i}sica, Universidad de La Laguna, E-38206 La Laguna, Tenerife, Spain
\label{ULL}
}

\abstract{%
We derive the oxygen abundance (O/H), the nitrogen-to-oxygen (N/O) abundance
ratio, and their corresponding radial gradients for a sample of 1431 galaxies
from MaNGA DR15 survey using two different realizations of the strong line
method: empirical R calibration and the Bayesian model-based {\sc HII-CHI-mistry} 
({\sc HCm}) code.
We find that both abundance calculation methods reveal a correlation between
the O/H gradient and the stellar mass of a galaxy.
This relation is non-linear, with the steepest average gradients in the
intermediate mass range and flatter average gradients for high- and low-mass
galaxies.
The relation between the N/O gradient and the stellar mass is, on average,
non-linear with the steepest gradients in the intermediate mass range 
($\log(M/M_\sun) \sim 10$), flatter gradients for high-mass galaxies, 
and the flattest gradients for low-mass galaxies.
However, the general trend of steepening N/O gradient for higher masses,
reported in previous studies, remains evident.
We find a dependence between the O/H and N/O gradients and the galaxy mean 
stellar age traced by the $D$(4000) index.
For galaxies of lower masses, both gradients are, generally, steeper for
intermediate values of $D$(4000) and flatter for low and high values of
$D$(4000).
Only the most massive galaxies do not show this correlation.
We interpret this behaviour as an evolution of the metallicity gradients with
the age of stellar population.
Though the galaxies with a positive slope of the $D$(4000) radial gradient tend to
have a flatter O/H and N/O gradients, as compared to those with a negative
$D$(4000) gradient.
}

%%%%%%%%%%%%%%%%%%%%%%%%%%%%%%%%%%%%%%%%%%
%%%%%%%%%%%%%%%%%%%%%%%%%%%%%%%%%%%%%%%%%%

\titlerunning{O/H and N/O gradients vs. stellar age in MaNGA galaxies}
\authorrunning{Zinchenko et~al.}
\maketitle

%-------------------------------------------------------------------

\section{Introduction}

The chemical composition of galaxies plays an important role in their formation
and evolution during the lifetime of the Universe.
Hence, the oxygen abundance correlates with other integrated characteristics of
galaxies, as in the well-known case of the Mass--Metallicity Relation (MZR),
where the average oxygen abundance increases with stellar mass (or luminosity)
of a galaxy \citep{Searle1971,Shields1978,VilaCostas1992,Zaritsky1994,Thuan2010,
PerezMontero2013,Sanchez2014,Sanchez2019,Zinchenko2019a,Yates2020}.
In the individual disk galaxies, the oxygen abundance depends diversely on the
galactocentric distance, but at the present epoch and on average it decreases
with radius, forming the so-called radial metallicity gradient
\citep{Sanchez2012,Sanchez2014,Pilyugin2014a,SanchezMenguiano2016,Zinchenko2016,
Belfiore2017,SanchezMenguiano2018,Kreckel2019,Zinchenko2019a,Zinchenko2019b,
Zurita2021}.

Negative metallicity gradients can be explained by the inside-out growth of the
galactic disks \citep{Matteucci1989,Boissier1999,Chiappini2001}.
However, the chemical enrichment and, therefore, formation of the shape of the
radial metallicity distribution in a galaxy is a complex process that depends
on its star formation history (SFH), the gas inflows and outflows as a whole,
and also on the different evolution timescales of its local parts
\citep{Troncoso2014,Zahid2014,Bothwell2016}.
This has two important implications regarding the formation of metallicity
gradients in disk galaxies.
First, galaxies with different histories should show different slopes of the
radial metallicity distribution.
From the observational point of view this is revealed in a large measured
scatter of metallicity gradient slopes, which is reported in many works
\citep[e.g.,][]{Sanchez2014,PerezMontero2016,Zinchenko2016,Belfiore2017}.
Second, radial metallicity gradients depend on integrated properties of
galaxies that reflect their evolution.

The most obvious parameter that defines the evolution of a galaxy is its mass.
Some studies suggest that the oxygen abundance gradient may depend on the
stellar mass of a galaxy \citep{Ho2015,Belfiore2017,Carton2018,Zinchenko2019b},
while other works do not report such a correlation
\citep{Sanchez2014,PerezMontero2016,SanchezMenguiano2016,SanchezMenguiano2018}.

A correlation between the global (average) gas-phase oxygen abundance and the $D$(4000)
index, which is an indicator of stellar age, has been found by \citet{Lian2015}
and \citet{SanchezMenguiano2020}.
Moreover, \citet{SanchezMenguiano2020} reported a positive correlation for the
local oxygen abundance as well.

Negative nitrogen-to-oxygen (N/O) gradients have been reported for a majority 
of galaxies \citep{Pilyugin2004,PerezMontero2016,Belfiore2017}. 
The mass--N/O relation has been studied for the first time by
\citet{PerezMontero2009b}, who have shown that the N/O ratio is larger at
higher stellar masses, similarly to the case of the MZR for oxygen. 
Theoretical models suggest that other variables, such as the time delay 
in nitrogen enrichment, 
the infall time scale, or the star formation efficiency could affect 
N/O ratio \citep{Molla2006}.
A possible correlation between the N/O gradient and the stellar mass has been
studied by \citet{PerezMontero2016} for 350 spiral galaxies from the CALIFA
survey.
They did not find a statistically significant correlation between the N/O
gradient and the stellar mass.
Later, for a larger sample of 550 galaxies from the MaNGA survey,
\cite{Belfiore2017} reported a steepening of the N/O gradient with the stellar
mass.

In this work, we study the distributions of the oxygen abundance and the N/O
abundance ratio for a large sample of 1431 nearby galaxies from the Mapping
Nearby Galaxies at Apache Point Observatory \citep[MaNGA;][]{Bundy2015} survey
data release 15 (DR15), which is a part of the Sloan Digital Sky Survey IV
\citep[SDSS IV;][]{Blanton2017}.
Our main goal is to explore the connection between the gas-phase metallicity
gradients and the stellar age of a galaxy.
We consider oxygen abundance gradients as well as N/O gradients and their
connection with the stellar mass.
The robust estimation of the chemical abundance is crucial for studying such
relations, therefore we take advantage of two precise strong line methods, which
are based on different observational and theoretical sets of calibration data.

The paper is structured as follows. In Section~\ref{sect:data} we describe 
the data, the sample selection criteria, and the spectral fitting procedure.
In Section~\ref{sect:abund} we discuss the methods used for determining O/H,
N/O, and the corresponding radial gradients in our sample, and we explore how
the abundance gradients are connected with the stellar age.
Finally, in Section~\ref{section:Summary} we summarize the main results of our
work.

%%%%%%%%%%%%%%%%%%%%%%%%%%%%%%%%%%%%%%%%%%%%%%%%
\section{Data}
\label{sect:data}
%%%%%%%%%%%%%%%%%%%%%%%%%%%%%%%%%%%%%%%%%%%%%%%%

\begin{figure}
\resizebox{1.00\hsize}{!}{\includegraphics[angle=000]{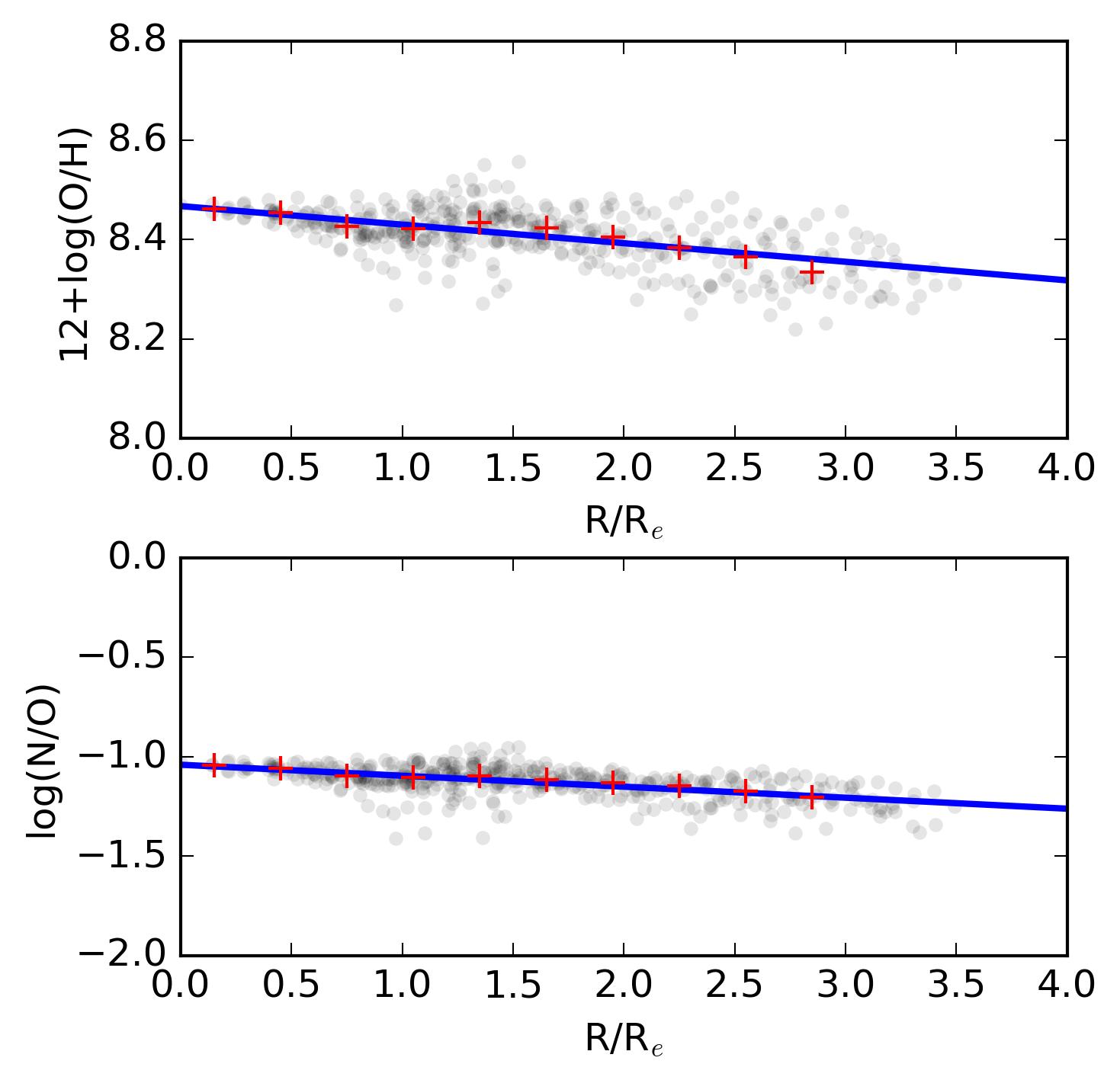}}
\caption{%
    An example of the radial gradients of O/H (upper panel) and N/O
    (lower panel) in the MaNGA galaxy 1-48157 using data from data cube
    10001-12701.
    Gray points are values in each spaxel, derived from the empirical R
    calibration of \citet{PilyuginGrebel2016}.
    Red crosses are median values in bins.
    A solid line is the linear fit to the data.%
}
\label{figure:grad-example}
\end{figure}

\begin{figure}
\resizebox{0.97\hsize}{!}{\includegraphics[angle=000]{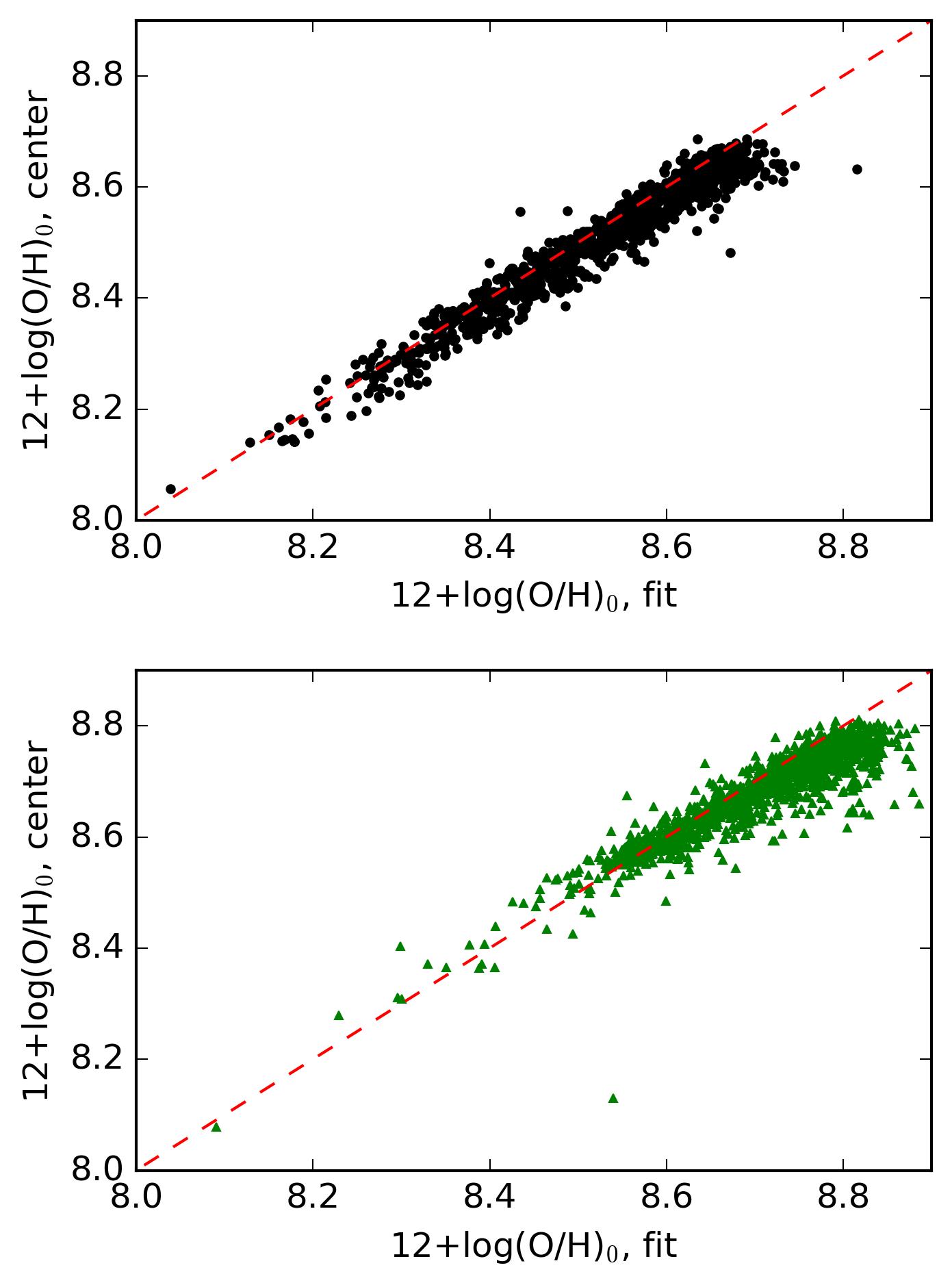}}
\caption{%
    Comparison between O/H at galaxy centers derived by extrapolating 
    the radial fits and by averaging in individual spaxels at $R < 0.1R_\text{e}$.
    Upper panel: O/H derived from the empirical R calibration. 
    Lower panel: O/H derived using the {\sc HCm} code.
    Dashed lines represent the one-to-one correspondence.%
}
\label{figure:manga-center-compare}
\end{figure}

\begin{figure}
\resizebox{1.00\hsize}{!}{\includegraphics[angle=000]{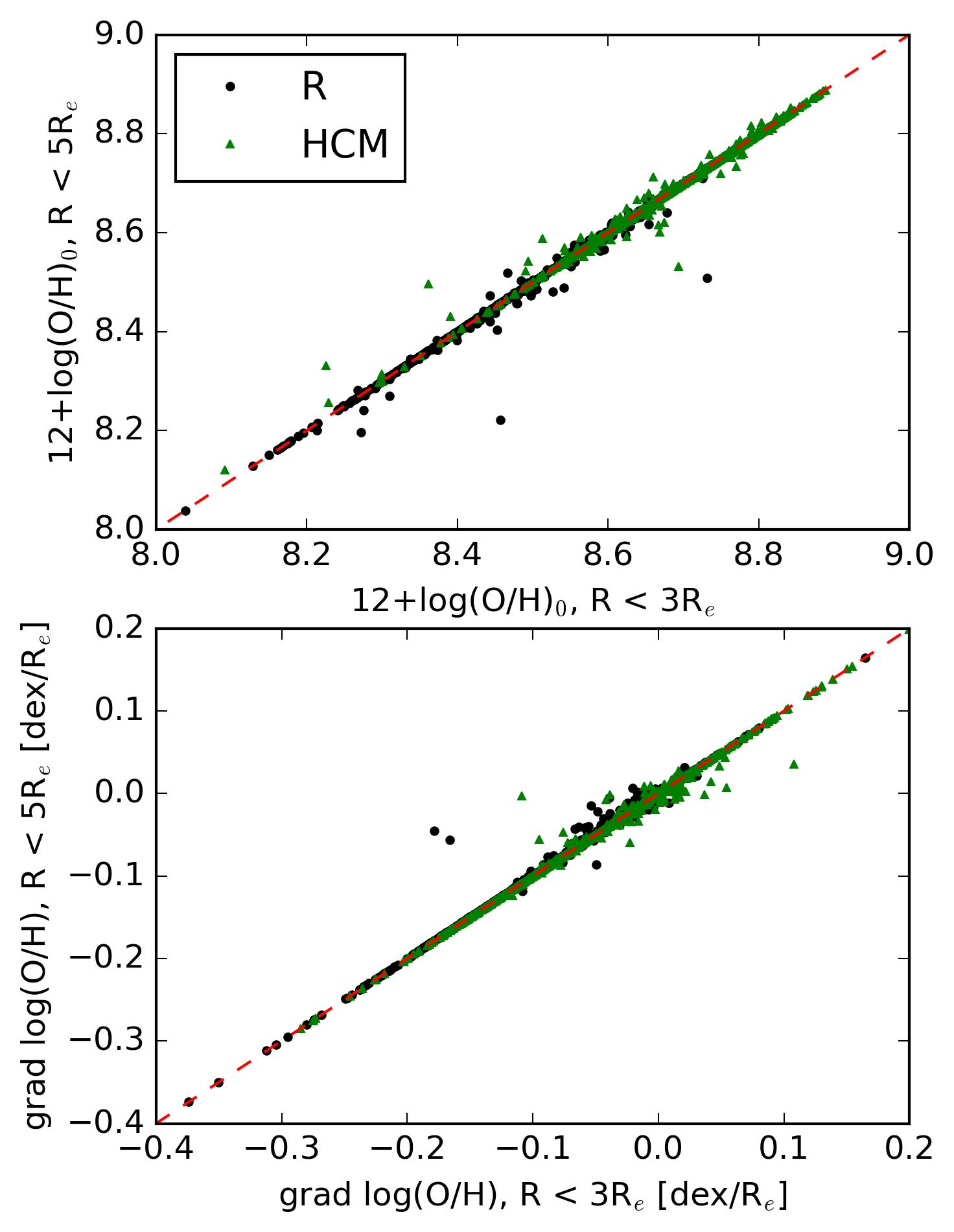}}
\caption{%
    Comparison between the fits to the derived O/H radial distributions at a
    galactocentric radius less than $3R_\text{e}$ (horizontal axes) and less than
    $5R_\text{e}$ (vertical axes).
    Upper panel: comparison of the extrapolated O/H abundances in the galaxy centers.
    Lower panel: comparison of the slopes of the fits.
    The oxygen abundances and their corresponding gradients have been derived
    using the two methods: the R calibration (black circles) or the {\sc HCm}
    code (green triangles).
    Dashed lines represent the one-to-one correspondence.%
}
\label{figure:fit-compare}
\end{figure}

For this study we prepared a sample of galaxies from the MaNGA SDSS DR15 survey
\citep{Bundy2015}.
We analysed the MaNGA spectra following \citet{Zinchenko2016}.
In brief, the stellar background in all spaxels is fitted using the public
version of the STARLIGHT code \citep{CidFernandes2005,Mateus2006,Asari2007}
adapted for execution in the NorduGrid ARC%
\footnote{\href{http://www.nordugrid.org/}{www.nordugrid.org}}
environment of the Ukrainian National Grid.
To fit the stellar spectra we used simple stellar population (SSP) spectra from
the evolutionary synthesis models by \citet{BC03} with ages from 1~Myr up to
13~Gyr and metallicities $Z$ of 0.004, 0.02, and 0.05.
The resulting stellar spectrum is subtracted from the observed spectrum to 
obtain a pure gas spectrum.
The line intensities in the gas spectrum were measured by fitting single
Gaussian profiles on the pure emission spectra.

To fit the emission lines we used our code ELF3D for emission line fitting in
the optical spectra.
The code is based on the \textit{iminuit} library \citep{iminuit}, which 
in turn is based on the SEAL Minuit2 code \citep{James1975}.
Since Minuit2 is sensitive to the choice of the initial parameters, we
implemented an option of the Monte Carlo (MC) approach to choose the initial
parameters of the fit. 
This approach significantly increases the robustness of line fluxes estimation.
The computational time per one data cube in the robust MC mode has been reduced
owing to the MPI parallelization of the code.
Our sample of galaxies was processed on two clusters, MareNostrum4 at the
Barcelona Supercomputing Center (BSC) and Golowood at the Main Astronomical 
Observatory of the NAS of Ukraine.

For each spectrum, we measure fluxes of the
[\ion{O}{II}]$\lambda\,\lambda$3727,3729,
H$\beta$,
[\ion{O}{III}]$\lambda$4959,
[\ion{O}{III}]$\lambda$5007,
[\ion{N}{II}]$\lambda$6548,
H$\alpha$,
[\ion{N}{II}]$\lambda$6584, and
[\ion{S}{II}]$\lambda$6717,6731 lines.
The line fluxes were corrected for interstellar reddening using the analytical
approximation of the Whitford interstellar reddening law \citep{Izotov1994},
assuming the Balmer line ratio of $\text{H}\alpha/\text{H}\beta = 2.86$.
When the measured value of $\text{H}\alpha/\text{H}\beta$ is less than 2.86,
the reddening is set to zero.

We apply the
$\log$([\ion{O}{III}]$\lambda$5007/H$\beta$) -- $\log$([\ion{N}{II}]$\lambda$6584/H$\alpha$) 
diagram \citep{BPT} and the dividing line proposed by \cite{Kauffmann2003} to
classify objects in two groups based on their main ionization source, which is
either massive stars or gas shocks and/or active galactic nuclei (AGNs).
We select only spectra with the signal-to-noise ratio $\text{SNR} > 5$ in all the
[\ion{O}{II}]$\lambda\,\lambda$3727,3729, H$\beta$, [\ion{O}{III}]$\lambda$5007,
H$\alpha$, [\ion{N}{II}]$\lambda$6584, and [\ion{S}{II}]$\lambda$6717,6731
lines.

Since the measurement of the radial metallicity gradient in the distant and
edge-on galaxies may be affected by significant systematical and statistical
errors, we only selected galaxies at distances less than 300~Mpc, which
corresponds to $z \sim 0.07$, and with the minor-to-major axis ratio of
$b/a > 0.35$ or an inclination less than $i \sim 70^\circ$.

Stellar masses, effective radii, and colors for the MaNGA sample were taken 
from the NASA-Sloan Atlas (NSA) catalog\footnote{\href{http://nsatlas.org}{nsatlas.org}}.
Stellar masses were derived from the K-correction fit for elliptical Petrosian
fluxes using the \citet{Chabrier2003} initial mass function and simple stellar
population models from \citet{BC03}.
Effective radii are defined as Sersic 50\% light radius along the major axis in
the $r$ band.

The inclination and the position angle of the major axis of the galaxies have
been obtained from the Sersic fit to the surface brightness profile in the
$r$~band available in the NSA catalog.

%%%%%%%%%%%%%%%%%%%%%%%%%%%%%%%%%%%%%%%%%%%%%%%%
\section{The gradients of O/H and N/O abundance ratios}
\label{sect:abund}
%%%%%%%%%%%%%%%%%%%%%%%%%%%%%%%%%%%%%%%%%%%%%%%%

\begin{figure*}
\resizebox{1.00\hsize}{!}{\includegraphics[angle=000]{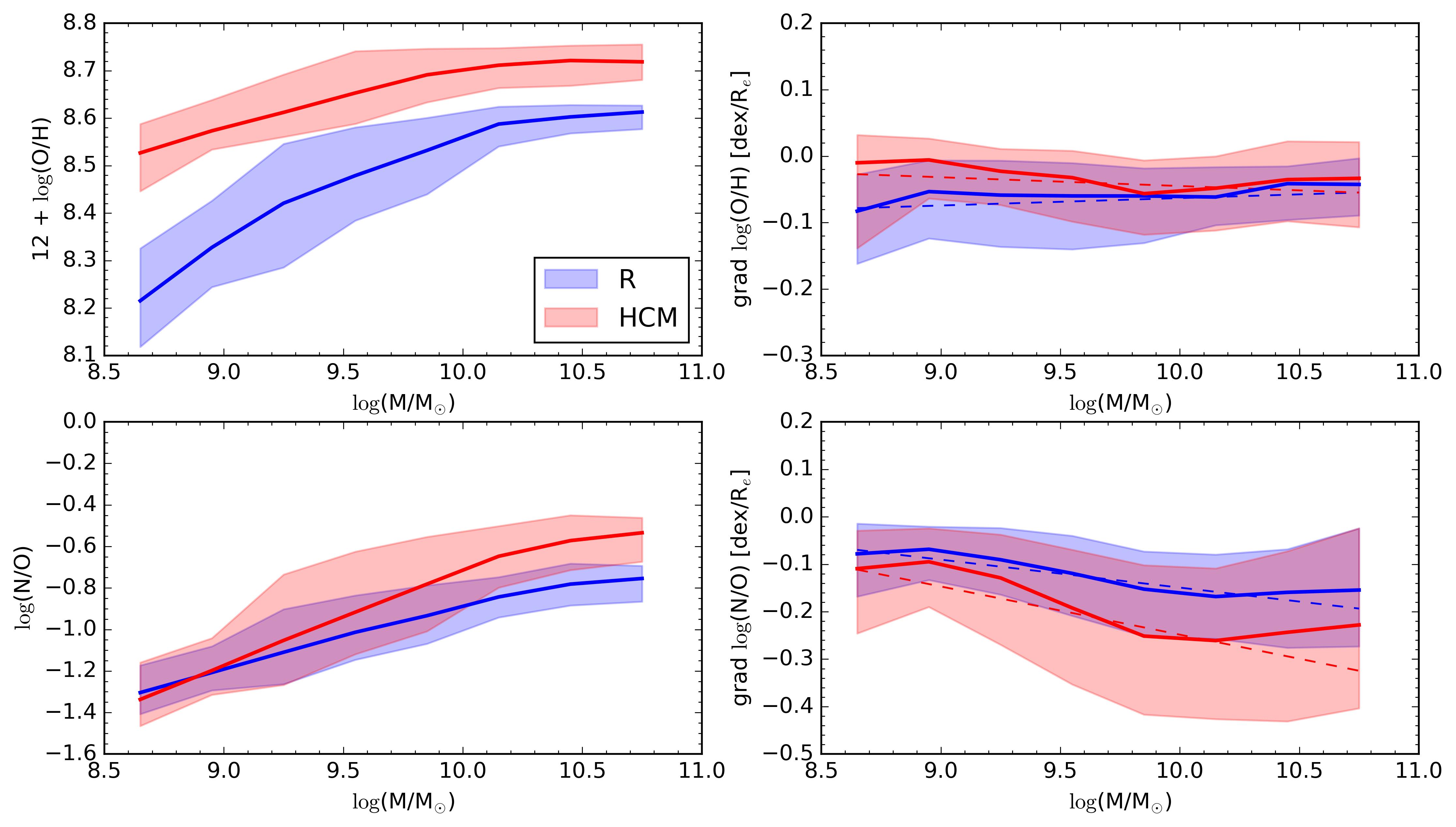}}
\caption{%
    The galaxy-center oxygen abundance $\log(\text{O}/\text{H})$, the
    nitrogen-to-oxygen ratio $\log(\text{N}/\text{O})$, and their gradients as
    a function of galaxy mass $\log(M/M_\sun)$.
    Median values (solid lines) with 1-$\sigma$ ranges scatter (shaded areas) 
    are shown in each mass bin.
    The metallicity parameters were calculated using the two methods: the R
    calibration (blue lines and areas) or the {\sc HCm} method (red lines and
    areas).
    For the O/H and N/O gradients, dashed lines show linear fits as a function
    of the stellar mass.%
}
\label{figure:MZ}
\end{figure*}

\begin{figure}
\resizebox{1.00\hsize}{!}{\includegraphics[angle=000]{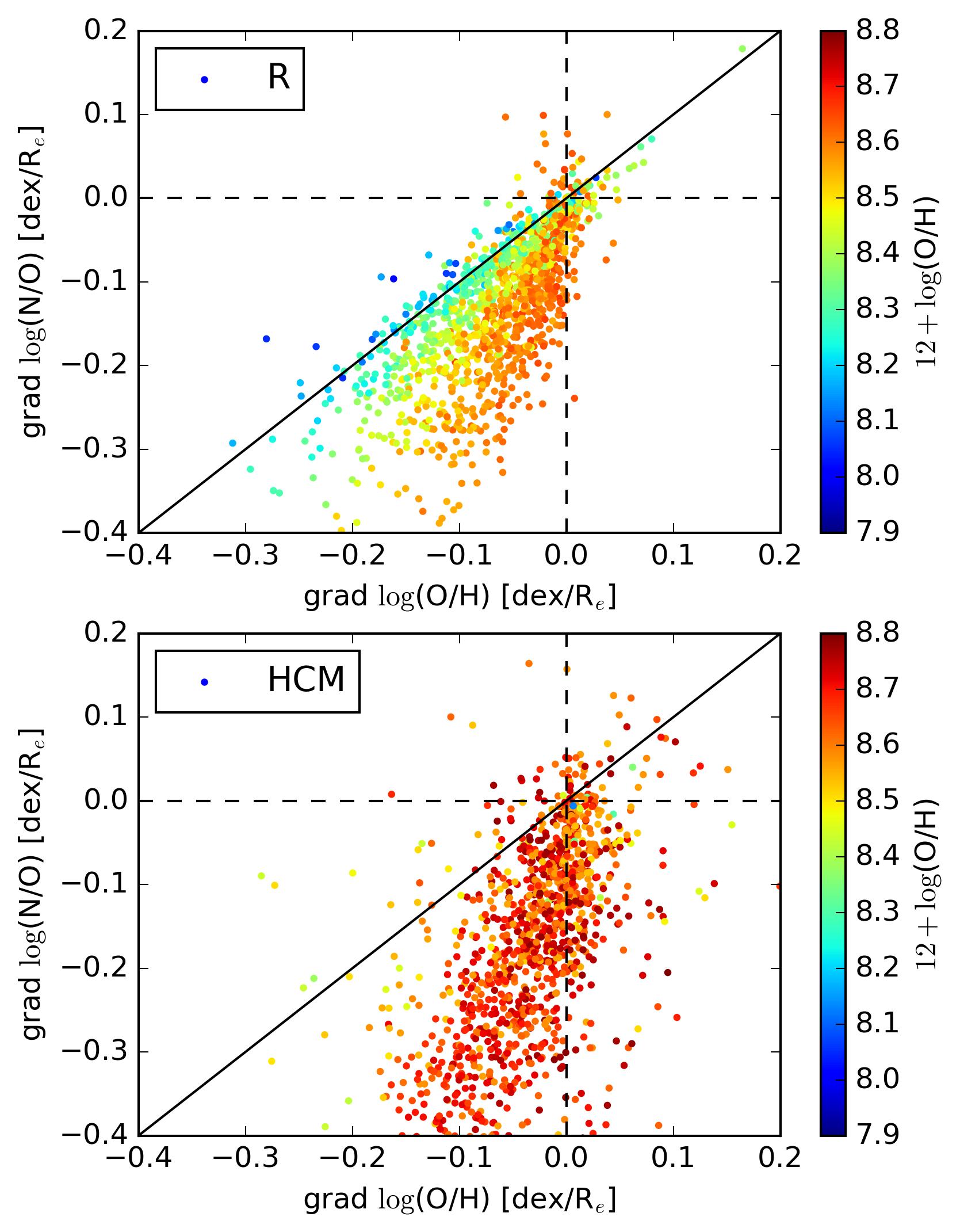}}
\caption{%
    Comparison of the oxygen abundance $\log(\text{O}/\text{H})$ gradient and
    the $\log(\text{N}/\text{O})$ ratio gradient for metallicities obtained from
    the R calibration (top panel) or the {\sc HCm} method (bottom panel).
    Color encodes the oxygen abundance at $R_\text{e}$.
    Black dashed are lines of zero gradients.
    Solid black line is a one-to-one correspondence for reference.%
}
\label{figure:grad-ON-NO}
\end{figure}

\subsection{Abundance determination}

The oxygen abundance $12 + \log(\text{O}/\text{H})$ and the nitrogen-to-oxygen
ratio $\log(\text{N}/\text{O})$ were derived using the empirical R calibration
\citep{PilyuginGrebel2016} and the Bayesian model-based code
{\sc HII-CHI-mistry} ({\sc HCm})%
\footnote{\href{https://www.iaa.csic.es/~epm/HII-CHI-mistry-opt.html}{www.iaa.csic.es/{\textasciitilde}epm/HII-CHI-mistry-opt.html}}
(hereinafter HCm) \citep{PM2014}.
We used {\sc HCm} version 4.1, assuming photoionization models from Cloudy
v.17.00 \citep{Ferland2017} with SEDs from POPSTAR \citep{Molla2009}, with
linear interpolations on the grid of O/H, N/O.
The R calibration for the oxygen abundance determination is divided in two
branches, depending on the metallicity range,
\begin{multline}
    (\mathrm{O/H})_\mathrm{U}
    = 8.589 + 0.022\,\log(R_3 / R_2) + 0.399\,\log N_2 \\
      + \Bigl(
            -0.137 + 0.164\,\log(R_3 / R_2) + 0.589\,\log N_2
        \Bigr) \log R_2
    \label{equation:ohru}
\end{multline}
for \ion{H}{II} regions with $\log N_2 \ge -0.6$ (the upper branch), and
\begin{multline}
    (\mathrm{O/H})_\mathrm{L}
    = 7.932 + 0.944\,\log(R_3 / R_2) + 0.695\,\log N_2 \\
      + \Bigl(
            0.970 - 0.291\,\log(R_3 / R_2) - 0.019\,\log N_2
        \Bigr) \log R_2
    \label{equation:ohrl}
\end{multline}
for $\log N_2 < -0.6$ (the lower branch).
Here $(\mathrm{O/H})_\mathrm{U}$ and $(\mathrm{O/H})_\mathrm{L}$ correspond to
$12 + \log(\text{O}/\text{H})$ for the upper and lower branches,
respectively.
For the determination of the N/O ratio we used the following calibration from
\citet{PilyuginGrebel2016} as well:
\begin{multline}
    \log(\mathrm{N/O})
    = -0.657 - 0.201\,\log N_2 \\
        + \Bigl( 0.742 - 0.075\,\log N_2 \Bigr) \log(N_2 / R_2).
    \label{equation:nolin}
\end{multline}

For the sake of consistency, in both strong line methods for 
the oxygen abundance determination we used the following set of emission lines:
[\ion{O}{II}]$\lambda\,\lambda$3727,3729, [\ion{O}{III}]$\lambda$5007, and
[\ion{N}{II}]$\lambda$6584.
Taking into account the [\ion{N}{II}]$\lambda$6584 line allows us to solve the
double-valued problem, which is typical for calibrations based on metallicity
indicators with oxygen lines only.
The downside of this approach is that oxygen abundance indicators based on the
[\ion{N}{II}]$\lambda$6584 line may introduce the N/O dependence into the
calibration relationship \citep{Schaefer2020}.
To overcome this problem the {\sc HCm} code uses a grid of models with different
N/O ratio.
The R calibration adopts a three-dimensional relation that incorporates a
correction for the N/O ratio by introducing indexes sensitive to the N/O ratio.

We note that the {\sc HCm} code allows us to use the
[\ion{S}{II}]$\lambda\,\lambda$6717,6731 lines in addition to those three lines
and that there is the S calibration proposed by \citet{PilyuginGrebel2016},
which replaces the [\ion{O}{II}]$\lambda\,\lambda$3727,3729 lines with the
[\ion{S}{II}]$\lambda\,\lambda$6717,6731 lines.
However, S$^+$ has a lower ionization potential compared to O$^+$ and,
therefore, could be more contaminated by the diffuse ionized gas (DIG).
Since MaNGA spectra have the [\ion{O}{II}]$\lambda\,\lambda$3727,3729 lines
starting from the rest frame, we prefer not to use the additional
[\ion{S}{II}]$\lambda\,\lambda$6717,6731 lines for the abundance determination.

We calculate the radial gradients for both $12 + \log(\text{O}/\text{H})$ and
$\log(\text{N}/\text{O})$ only for galaxies with at least 10~spaxels with 
measured $12 + \log(\text{O}/\text{H})$ and $\log(\text{N}/\text{O})$ values.
To make fits more robust we require that data points should be well spread 
along the galactocentric distance, i.e. the galactocentric distance of the
outermost data point should be at least $1 R_\text{e}$ higher compared to the
distance of the innermost data point.
To obtain the radial gradient and its error, a least square linear fit has been
performed on 5\,000 bootstrap samples with replacement.
In the end, we selected 1\,431 galaxies for further analysis.
Figure~\ref{figure:grad-example} illustrates an example of the radial gradients
of $12 + \log(\text{O}/\text{H})$ and $\log(\text{N}/\text{O})$ for one of the
selected galaxies.

Since the radial distribution of the oxygen abundance may be more complex than
the pure linear distribution \citep{Vilchez1988c,Bresolin2009,Goddard2011,RosalesOrtega2011,
SanchezMenguiano2016,Belfiore2017,SanchezMenguiano2018,Pilyugin2017},
in Figure~\ref{figure:manga-center-compare} we compare the oxygen abundances in
the galaxy centers derived either by extrapolating the fits or by averaging in
individual spaxels at $R < 0.1R_\text{e}$.
We see that the oxygen abundances in the centers of high-metallicity galaxies
are lower when averaged over the innermost spaxels then when extrapolated from
the fits, which is true both for the empirical and theoretical methods used.
The abundances show the same behavior when derived by either of the both
methods.
Thus, we reaffirm the conclusion of previous studies, which have found a
flattening of the metallicity gradient in the central parts of massive galaxies
\citep{Belfiore2017,SanchezMenguiano2018,Zinchenko2019a}.
We see the same behavior for the N/O ratio: in the centers of high-metallicity
galaxies it is generally lower when averaged rather when it is extrapolated.

The other widely reported feature of the radial metallicity distribution is its
flattening in the outer parts of the disk \citep{Bresolin2009,Goddard2011,
SanchezMenguiano2016,Belfiore2017,SanchezMenguiano2018}.
We checked for the existence of this outer flattening and its possible effect
on the calculated gradients.
In Figure~\ref{figure:fit-compare} we compare the oxygen abundance and its
gradient derived in a galactocentric radius either within $3R_\text{e}$ or $5R_\text{e}$.
However, it should be noted that the fraction of metallicity measurements in the outer 
part of galaxies is low, about 1\% for $R/R_\text{e} > 3$ and 0.2\% for $R/R_\text{e} > 5$.
These two cases show no systematic difference neither in the oxygen abundance
extrapolated to the center of a galaxy nor in the slope of the fit.
Neither the N/O gradients show any dependence on the maximum galactocentric
radius adopted for the calculation of the gradients.
Therefore, for further analysis we use the gradients derived within $5R_\text{e}$.

\subsection{Abundances and gradients as a function of mass}

In Figure~\ref{figure:MZ} we show relations between the parameters of the
derived linear radial fits for both O/H and N/O as a function of the stellar
mass in the galaxies from our sample.

We show the oxygen abundance at $R_\text{e}$ as a function of the stellar mass in the
top left panel of Fig.~\ref{figure:MZ}.
Both methods produce different abundances that differ by ${\sim}0.1$~dex for
massive galaxies and this difference grows to ${\sim}0.3$~dex in the low-mass
regime.
But at the same time the oxygen abundance steeply increases up to stellar masses
of some $10^{10} M_\sun$ and then it stalls for higher masses.
This flattening at high masses is in agreement with many previous studies
\citep[e.g.,][]{Tremonti2004,Kewley2008,Thuan2010,Sanchez2019,Zinchenko2019a}.
The MZR relation in Fig.~\ref{figure:MZ}, obtained with the {\sc HCm}-based O/H
abundances, reproduces well the one obtained by \citet{PerezMontero2016} for the
CALIFA sample using the same {\sc HCm} code.
In both cases, the median O/H abundance increases with the stellar mass from
${\sim}8.5$~dex to ${\sim}8.7$~dex having a flattening at high masses.
This implies only ${\sim}0.2$~dex change in O/H across the MZR mass range,
$8.7 < \log(M/M_\sun) < 10.7$, what is significantly smaller compared to
0.4~dex in the case of the R calibration.
Other studies based on $R_{23}$ and O3N2 calibrations have also reported a
bigger change in O/H of ${\sim}$0.4--0.5~dex across a mass range of
$9 < \log(M/M_\sun) < 11$ \citep{Tremonti2004,Belfiore2017}.

The N/O ratio at $R_\text{e}$ also increases with the galaxy mass (see left bottom
panel  of Fig.~\ref{figure:MZ}).
There is no bias between the R and {\sc HCm} abundance ratios at low masses.
Meanwhile, at high masses, the N/O ratio derived by the {\sc HCm} method is on
average up~to 0.2~dex higher compared to the one derived using the R
calibration.
At the same time and compared to other studies, the R calibration yields the
same range of the N/O ratio, from $-1.3$~dex at low masses through $-0.8$~dex at
high masses, which is consistent with the results obtained by
\citet{Belfiore2017} using calibrations of the $R_{23}$ and O3N2 parameters and
\citet{PerezMontero2016} using the {\sc HCm} code version 2.0.

Both the R calibration and the {\sc HCm} methods confirm negative oxygen
abundance gradients for the majority of galaxies from our sample (see top right
panel of Fig.~\ref{figure:MZ}).
Median values of the oxygen abundance gradients are
$(-0.06\pm 0.06)~\text{dex}/R_\text{e}$ for the R abundances and
$(-0.03\pm 0.06)~\text{dex}/R_\text{e}$ for the {\sc HCm} abundances.
We note that the distribution of gradients is not symmetric, it has a steeper
cutoff around zero and a longer tail at the lower end.
On average, the oxygen abundance gradients calculated using the R calibration
and the {\sc HCm} methods are consistent for massive galaxies.

The slope of the oxygen abundance gradient with respect to the stellar mass is
calculated by a linear fit to the data. The error of the linear fit is estimated 
by bootstrapping with 1000 random re-samplings of the data with replacement. 
For the R calibration the slope is mild positive while the {\sc HCm} method 
yields a mild negative slope.
Thus, the comparison between the linear fit and the O/H gradient averaged
in the mass bins shows that
both relations are significantly non-linear having a U-shape with
the steepest average gradients around $\log(M/M_\sun) = 10.0$ and flatter
average gradients in high- and low-mass bins.
These U-shape profiles are consistent with the result obtained by
\cite{Belfiore2017}, who used the $R_{23}$ parameter as calibrated by
\citet{Maiolino2008} and \citet{Poetrodjojo2021}, who compared a large set of strong line 
methods using 248 galaxies from the SAMI Galaxy Survey.
\citet{Zinchenko2019a} and \citet{Pilyugin2019} reported a mild positive slope
when applying the R calibration to a smaller sample containing relatively
massive CALIFA and MaNGA galaxies.
Thus, undersampling of the low mass galaxies produces only a positive
correlation between the O/H gradient and the stellar mass for massive galaxies.
The same positive correlation between the O/H gradient and the stellar mass was
found by \citet{Tissera2019} for galaxies in the EAGLE simulation that have
quiet merger histories.

We show the average N/O gradients as a function of mass in the bottom right
panel of Fig.~\ref{figure:MZ}.
Compared to the O/H gradients, the N/O gradients are steeper with median and
standard deviation values of $(-0.12\pm 0.09)~\text{dex}/R_\text{e}$ for the R
abundances and $(-0.18\pm 0.16)~\text{dex}/R_\text{e}$ for the {\sc HCm} abundances.
On average, the N/O gradients correlate with the stellar mass too.
Both the R calibration and the {\sc HCm} method produce steeper average N/O
gradients for more massive galaxies, which is in tune with the previous result of
\citet{Belfiore2017}.
The R calibration provides a flatter slope of $(-0.06\pm 0.005)~\text{dex}/R_\text{e}$ 
per dex in stellar mass compared to $(-0.10\pm 0.01)~\text{dex}/R_\text{e}$ 
per dex in stellar mass obtained by the {\sc HCm} method.
However, considering this correlation in detail, we see a non-linear relation
between the average N/O gradient and the stellar mass, which has the steepest
slope in the intermediate mass range and some flattening in the high- and
low-mass ranges.

\subsection{Relation between O/H and N/O abundance gradients}

\begin{figure}
\resizebox{1.0\hsize}{!}{\includegraphics[angle=000]{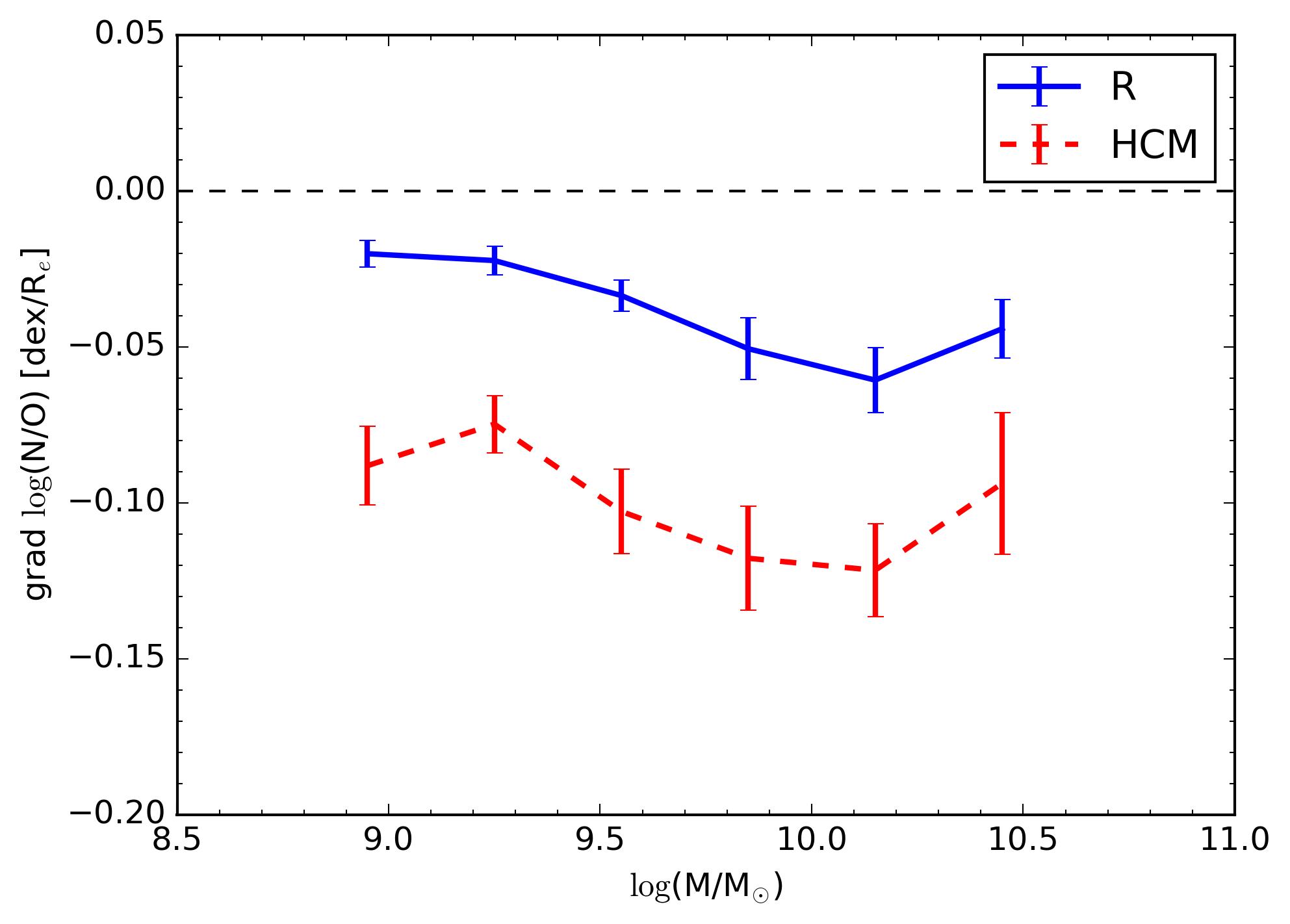}}
\caption{%
    The median N/O ratio gradient as a function of the galaxy mass for a sample
    of galaxies with a flat oxygen abundance gradient.
    The abundances are either derived from the R calibration (blue solid) or
    using the {\sc HCm} method (red dashed).
    Error bars show the standard deviation divided by the square root of the
    number of data points in each mass bin.
    For reference, a horizontal dashed line shows the zero level of the N/O
    ratio gradient.%
}
\label{figure:grad-NO-zeroO}
\end{figure}

\begin{figure*}
\resizebox{1.01\hsize}{!}{\includegraphics[angle=000]{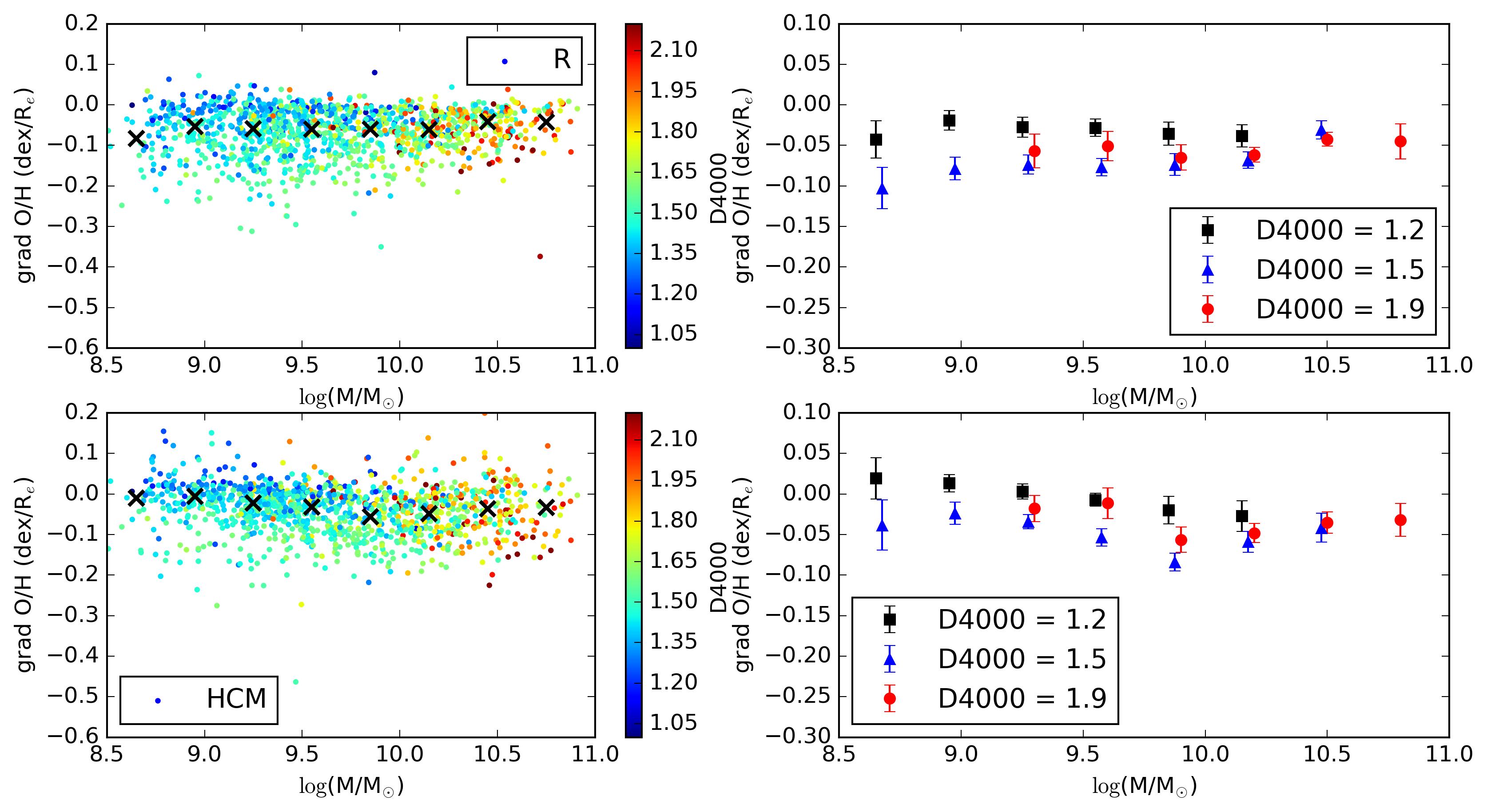}}
\caption{%
    The oxygen abundance gradient as a function of the stellar mass and the
    $D(4000)$ index for abundances derived from either the R calibration
    (top panels) or using the {\sc HCm} method (bottom panels).
    Left panels: the O/H gradient as a function of the stellar mass of a galaxy
    with color-coded $D(4000)$ for individual galaxies; crosses are median
    values in each mass bin.
    Right panels: the median O/H gradient with 2$\sigma$~CI in each mass bin for
    three subsamples of galaxies with different ranges of the $D(4000)$ index:
    1.0--1.4 (black), 1.4--1.6 (blue), 1.6--2.2 (red).%\\\hspace{\textwidth}%
}
\label{figure:MZparamOH-D4000}
\end{figure*}

\begin{figure*}[htb!]
\resizebox{1.01\hsize}{!}{\includegraphics[angle=000]{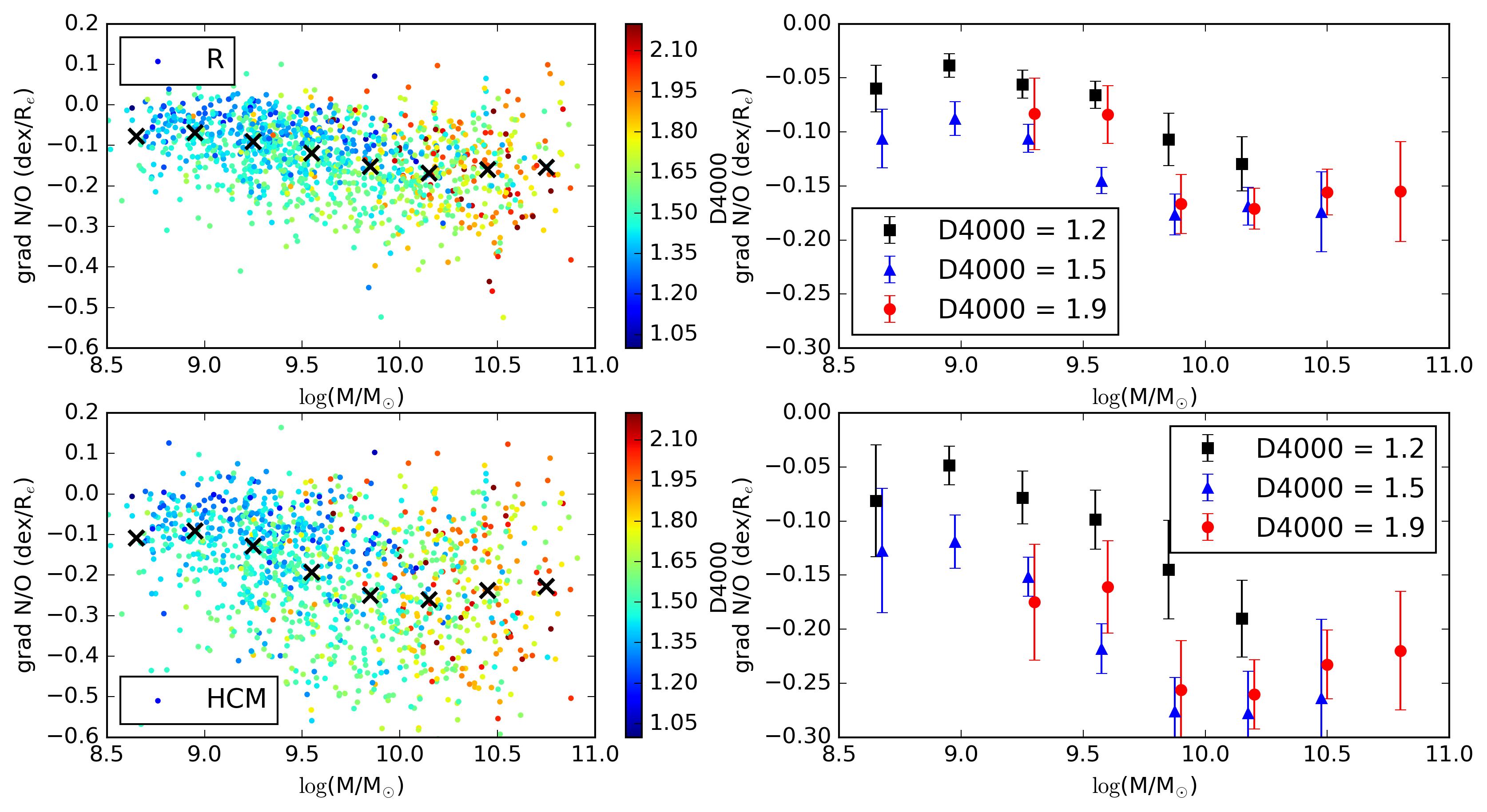}}
\caption{%
    Same as in Fig.~\ref{figure:MZparamOH-D4000} but for the N/O gradient as a
    function of the stellar mass and the $D(4000)$ index.
    The layout of panels, the notations, and the three subsamples of $D(4000)$
    are the same.%
}
\label{figure:MZparamNO-D4000}
\end{figure*}

In the previous subsection, we showed that the median N/O gradient correlates
with the stellar mass of a galaxy.
As it has been suggested by \citet{PerezMontero2009b}, this can be related not to
the galaxy mass itself but to a correlation between the galaxy mass and its
oxygen abundance (the mass--metallicity relation).
In addition, at high metallicity, which is typical for the galaxies from our
sample, nitrogen is mostly produced as a secondary element in the CNO cycle,
what implies that a yield of N depends on the amount of C and O already present
in the star.
Therefore, the nitrogen production depends on metallicity and at high
metallicities the oxygen abundance and the N/O ratio are expected to have a
non-linear relation
\citep[see][among others]{VilaCostas1992,PerezMontero2013,PilyuginGrebel2016,
PerezMontero2016,Vincenzo2016}. 
Though inflow and outflow of gas may change how the O/H abundance and the N/O
ratio are related \citep{Edmunds1990,Koppen2005}.
These effects can be very significant at low metallicities
\citep{Amorin2010,Amorin2012}.
The variation of the star formation efficiency (SFE) can also affect O/H and N/O
ratios \citep{Molla2006,Molla2016}.
This mechanism is invoked as the probable cause for the observed N/O enhancement
in the centers of barred galaxies \citep{Florido2015A&A,Zurita2021}.

If the N/O ratio depends on the oxygen abundance one can expect a correlation
between the N/O gradient and the O/H gradient as well.
Moreover, this correlation should have particular features.
First, the N/O radial distribution should be flat if the O/H radial distribution
is flat.
Second, both gradients should be equal at low metallicity where nitrogen is
mostly produced as a primary element and the N/O ratio is constant.
At higher metallicity, when nitrogen is mostly produced as a secondary element,
the N/O ratio gradient should be steeper compared to the O/H one.

All three features are seen in Figure~\ref{figure:grad-ON-NO} for 
the R abundances (top panel). In the case of the abundances derived by the {\sc HCm} method,
the N/O gradients also increase with O/H gradients in the case of 
the galaxies with high metallicity (bottom panel in Figure~\ref{figure:grad-ON-NO}). 
Nevertheless, in the case of the {\sc HCm} method, as it can be seen in top-left panel 
of Figure~\ref{figure:MZ}, almost all points lie above the low metallicity regime 
(i.e. 12+log(O/H) > 8.4), so this can not be properly checked for this method.

In the previous paragraph, we explained that, on average, the slope of the N/O
gradient correlates with the slope of the O/H gradient as N/O correlates with
O/H.
However, this correlation does not exclude a simultaneous dependence of the
slope of the N/O gradient on the stellar mass.
To further investigate this, we represent in Fig.~\ref{figure:grad-NO-zeroO} 
the median N/O gradient as a function of the galaxy mass for a sample of
galaxies with flat oxygen abundance gradient (less than $0.02~\text{dex}/R_\text{e}$).
The N/O gradients for a sample of galaxies with flat O/H gradients show the same
behaviour with respect to the stellar mass as the ones in the full sample of
galaxies.
This confirms that the N/O gradient depends on the stellar mass independently of
the O/H gradient.

%%%%%%%%%%%%%%%%%%%%%%%%%%%%%%%%%%%%%%%%%%%%%%%%
\subsection{The relation between abundance gradients and stellar age}
%%%%%%%%%%%%%%%%%%%%%%%%%%%%%%%%%%%%%%%%%%%%%%%%

\begin{figure}
\resizebox{1.00\hsize}{!}{\includegraphics[angle=000]{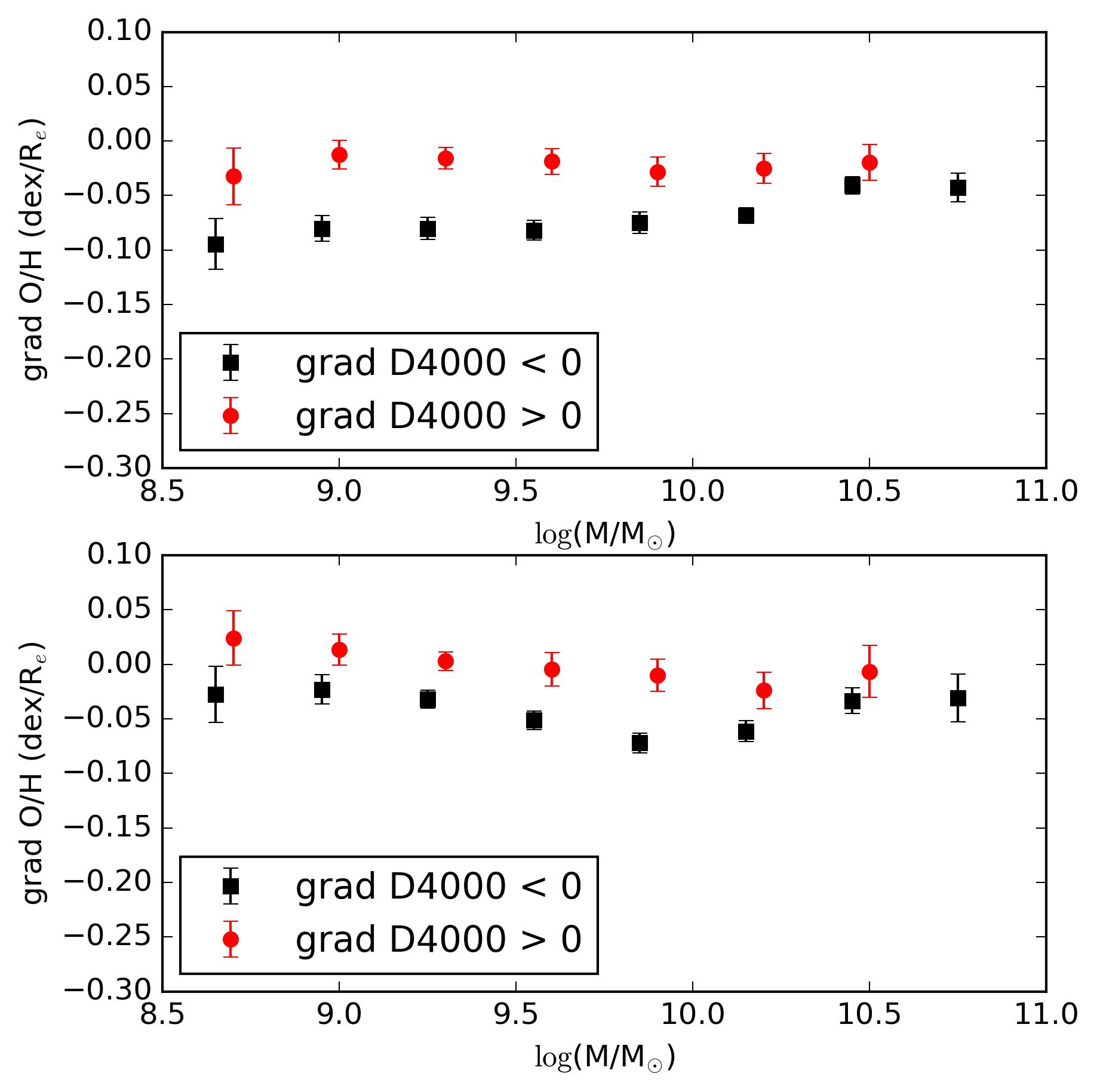}}
\caption{%
    The oxygen abundance gradient as a function of the stellar mass and the
    gradient of the $D(4000)$ index for abundances derived from either the R
    calibration (top panel) or using the {\sc HCm} method (bottom panel).
    Shown are median O/H gradients with 2$\sigma$~CI in each mass bin for two
    subsamples of galaxies with negative (black) and positive (red) gradient of
    $D(4000)$.%
}
\label{figure:OH-D4000-grad}
\end{figure}

\begin{figure}
\resizebox{1.00\hsize}{!}{\includegraphics[angle=000]{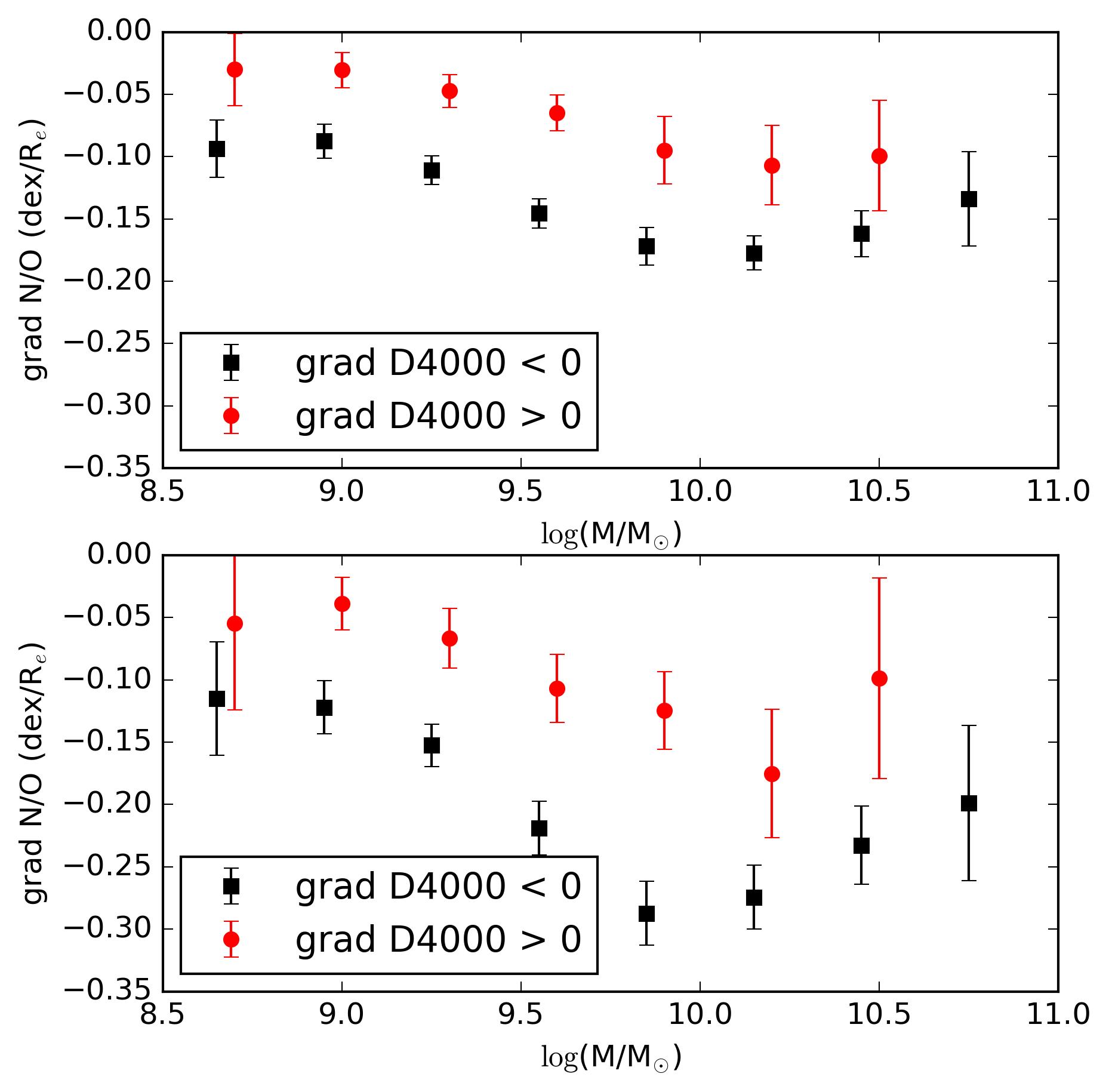}}
\caption{%
    Same as in Fig.~\ref{figure:OH-D4000-grad} but for the N/O gradient as a
    function of the stellar mass and the gradient of the $D(4000)$ index.
    The notations and the two subsamples of the $D(4000)$ gradient are the same.%
}
\label{figure:NO-D4000-grad}
\end{figure}

\begin{figure}
\resizebox{1.00\hsize}{!}{\includegraphics[angle=000]{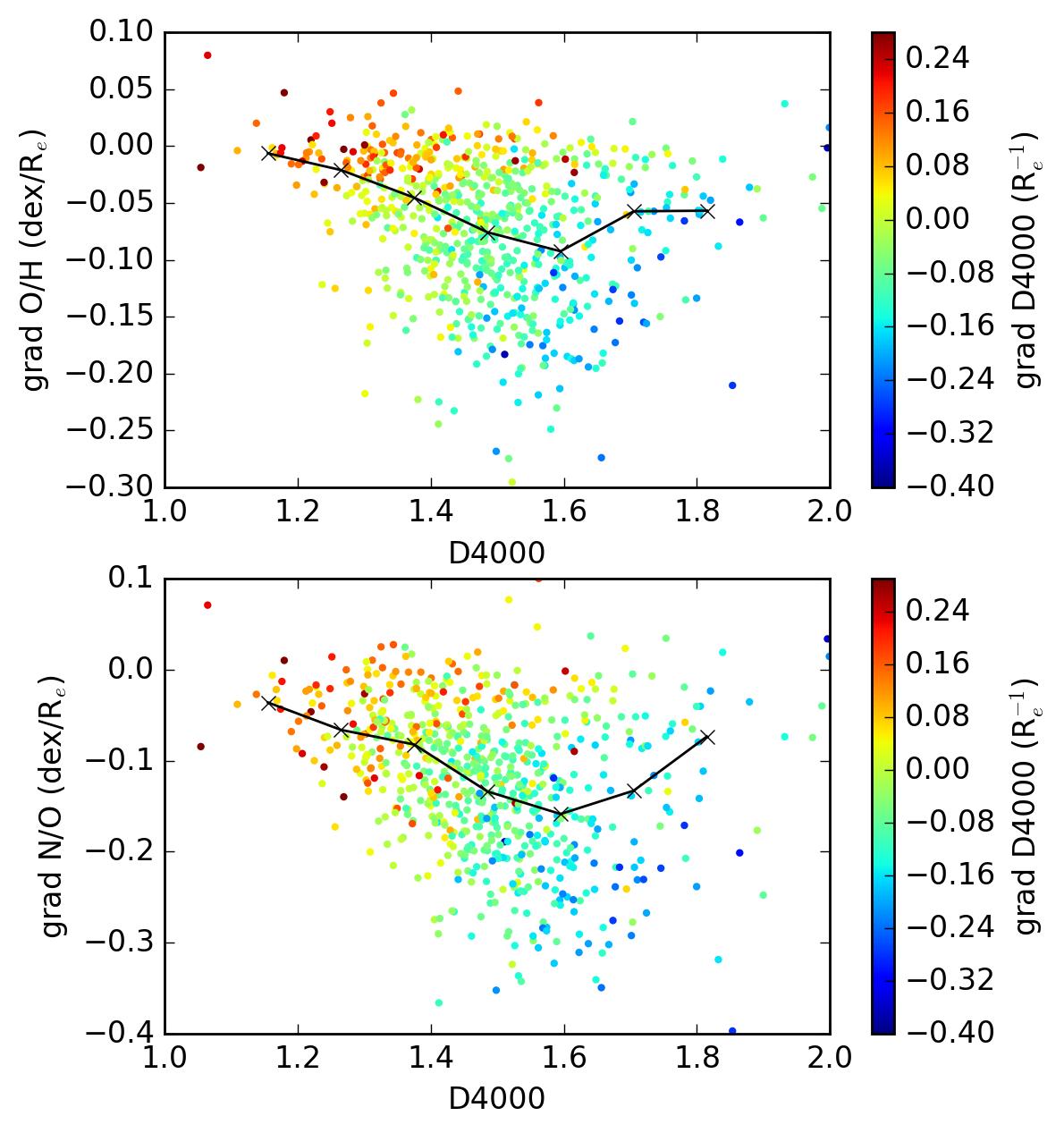}}
\caption{%
    The O/H gradient (top panel) and N/O gradient (bottom panel) as a function
    of the $D(4000)$ index with color-coded gradient of $D(4000)$ for abundances
    derived from the R calibration (see the text for details).
    Crosses connected by solid lines present median abundance gradient values in
    each bin of $D(4000)$.%
}
\label{figure:grad-3D}
\end{figure}

As it has been shown by \citet{Lian2015} and \citet{SanchezMenguiano2020}, the
FMR or the mass--metallicity relation may depend on the stellar age of the
galaxy.
These authors used the 4000-{\AA} break strength $D$(4000) as a proxy to the
stellar age \citep{Kauffmann2003b,Gallazzi2005,SanchezAlmeida2012}.
In this section, we discuss the links between the O/H and N/O gradients and the
average stellar age of the galaxy.
Since it is not easy to directly measure the stellar age, we adopted $D$(4000)
as a proxy to it. Another advantage of $D$(4000) is its sensitivity to the
stellar age variations during the last few Gyrs \citep{Poggianti1997,Noll2009},
which is enough to see its relation with the evolution of chemical abundances.

We derived values and gradients of the $D$(4000) index by fitting linearly the
radial distribution of the $D$(4000) 
index, calculated using the method of \citet{Balogh1999}, at Voronoi bins, taken from the
MaNGA Firefly value-added catalog \citep{Goddard2017,Parikh2018}.
To quantify the accuracy of the fit, we calculated the standard deviation
of the residuals of the fit to the individual $D$(4000) values, which is 0.07 for our
sample.
A visual inspection of the fit confirms that it is a good estimation of the
radial profile.
Although for some galaxies, a higher-order polynomial or segmented linear fit
may better represent the local values of $D$(4000), the estimation of the general
slope of $D$(4000) remains robust.
Further in this work, the $D$(4000) index of a galaxy is defined as a $D$(4000)
value from the fit at radius $R = 0$.

First, we investigate how stellar age affects O/H and N/O gradients.
Figure~\ref{figure:MZparamOH-D4000} shows the oxygen abundance gradient as a
function of the stellar mass and the $D(4000)$ index for abundances derived
from either the R calibration or using the {\sc HCm} method.
Left panels present the O/H gradient against the mass as distributed in
individual galaxies with color-coded $D(4000)$ index.
Right panels show the median O/H gradient in mass bins for three subsamples of
galaxies with $1.0 < D(4000) < 1.4$, $1.4 < D(4000) < 1.6$, and
$1.6 < D(4000) < 2.2$, with a similar number of galaxies in each of them.
Since the $D(4000)$ value increases with the average stellar age, we refer to
these three as to young, intermediate, and old subsamples.
Error bars show 2$\sigma$ confidence intervals for the median O/H gradients
where $\sigma$ is calculated in each bin as the standard deviation divided by
the square root of the number of data points.
Figure~\ref{figure:MZparamNO-D4000} presents the same diagram for the radial
gradient of the N/O ratio.

In the case of the oxygen abundance gradient, the same trend results from either
the R calibration or by using the {\sc HCm} method.
For galaxies with $\log(M/M_\sun) < 10.25$ the oxygen abundance gradient differs
depending on the $D(4000)$ value, it is steeper in the intermediate subsample
compared to the young and old subsamples.
We see the same trend for the N/O gradient derived either from the R calibration
or by using the {\sc HCm} method.

Although galaxies with low $D(4000)$ in our sample are not the progenitors
of galaxies with higher $D(4000)$, this index characterizes how old is the
stellar population of a galaxy. Therefore, if the galaxies in our sample were not
simultaneously formed, then $D(4000)$ provides us an indirect opportunity to
trace their evolution. Bearing this in mind, 
the trends seen on these plots can be interpreted as an evolution of the
gradients of oxygen and the N/O ratio with time.
In the beginning, galaxies with young stellar populations develop shallow oxygen
abundance gradients and likely shallow N/O ratio gradients as well.
Then, with the aging of the stellar population the gradients become steeper.
This may be caused by faster chemical evolution of the inner and denser parts of
the galaxies.
However, the chemical enrichment is limited by the interstellar gas
availability, so when a significant fraction of it has been consumed to form new
stars in the central part of galaxies, the pace of the chemical enrichment in
the outer parts becomes relatively faster.
This leads to the flattening of the radial gradients of chemical elements in the
galaxies with an old stellar population.
In the current epoch, this trend can be seen as a flattening of the radial
metallicity gradient in the most massive galaxies
\citep{Belfiore2017,SanchezMenguiano2018}, which have at the same time old
stellar populations and a very limited amount of interstellar gas.
This can explain why our sample of galaxies does not show any correlation of the
O/H and N/O gradients with the $D(4000)$ index at high stellar masses.
It is important to note that such trends are predicted by the inside-out growth 
scenario of the galactic disks \citep{Matteucci1989,Boissier1999,Chiappini2001}.

If the abundance gradient depends on the average stellar age, that is the
$D(4000)$ index of the galaxy, one could also expect that the abundance gradient
correlates with the radial gradient of $D(4000)$.
In Figure~\ref{figure:OH-D4000-grad} we show the oxygen abundance gradient with
respect to the stellar mass of a galaxy for galaxies with positive and negative
radial gradient of $D(4000)$.
On average, galaxies with the positive gradient of $D(4000)$ show significantly
flatter oxygen abundance gradients compared to the ones with the negative
gradient of $D(4000)$ in almost any mass bin for abundances obtained by either
of the both strong line methods.
As in the previous cases, the difference decreases for massive galaxies with
$\log(M/M_\sun) > 10.25$.

In Figure~\ref{figure:NO-D4000-grad} we present the same plot for N/O gradients.
As in the case of the oxygen abundance gradients, it is evident that the N/O
gradient flattens for galaxies with the positive gradient of $D(4000)$ up to
stellar masses of $\log(M/M_\sun) \approx 10.25$ for abundances derived using
either of the both strong line methods.

We have first demonstrated that O/H and N/O gradients correlate with
$D(4000)$ and then with the gradient of $D(4000)$ as well.
In Figure~\ref{figure:grad-3D} we explore these two correlations in detail for
the galaxies in a narrower stellar mass range of $9 < \log(M/M_\sun) < 10$, where 
the correlation of abundance gradients with $D(4000)$ is stronger 
and the range of $D(4000)$ values is wide.
In confirmation of the results described above in this section, both the O/H and
N/O gradients depend, on average, on the $D(4000)$ index and its gradient
simultaneously.
Another interesting behaviour, described above, is that the $D(4000)$ gradient
becomes steeper and negative when $D(4000)$ increases. Since here we consider
a relatively narrow range in stellar mass, we may see how stellar evolution along 
the discs correlates with the gradients of O/H and N/O. 

%%%%%%%%%%%%%%%%%%%%%%%%%%%%%%%%%%%%%%%%%%%%%%%%
\subsection{Comparison with theoretical models}
%%%%%%%%%%%%%%%%%%%%%%%%%%%%%%%%%%%%%%%%%%%%%%%%

Next, we compare our results with the predictions from theoretical models.
As we discussed above, the change of the radial metallicity gradient with
$D(4000)$ may be interpreted as its evolution with time. Indeed, a number of
theoretical models predicts such changes although there is a significant
disagreement in their conclusions regarding the evolution of the metallicity
gradient. For example, models of a Milky Way-like galaxy by \citet{Molla2019}
predict a slight flattening of the gradient starting from $z = 4$ with a
possible steepening from $z = 1$ to $z = 0$ in some models. An analytical model
developed by \citet{Belfiore2019} also predicts a general flattening of the
metallicity gradient with time, taking into account significant degeneracies
between model parameters at the present-day epoch.

In contrast, \citet{Sharda2021} predicted that the metallicity gradient in
massive galaxies has steepened over time until the advection-dominated regime
changed to the accretion-dominated regime, when the metallicity gradient started
to flatten. Assuming that $D(4000)$ correlates with the time passed from the
beginning of the active star formation in a galaxy, the relation presented in
Fig.~\ref{figure:grad-3D} can be explained by the \citet{Sharda2021} model.
Moreover, their Figure~12 shows that their model predicts a flattening of the
present-day metallicity gradient with the stellar mass. The same trend has been
observed in this work and in \citet{Belfiore2017} for massive galaxies in
the MaNGA survey as well as for galaxies in the CALIFA survey
\citep{Zinchenko2019b}.

%%%%%%%%%%%%%%%%%%%%%%%%%%%%%%%%%%%%%%%%%%%%%%%%
\section{Summary and conclusions}
\label{section:Summary}
%%%%%%%%%%%%%%%%%%%%%%%%%%%%%%%%%%%%%%%%%%%%%%%%

We have derived the oxygen abundance, $12 + \log(\text{O}/\text{H})$, the
nitrogen-to-oxygen abundance ratio, $\log(\text{N}/\text{O})$, and their
corresponding radial gradients normalized to the effective radius $R_\text{e}$ for a
sample of 1\,431 galaxies from the MaNGA DR15 survey using both the empirical R
calibration by \citet{PilyuginGrebel2016} and the model-based Bayesian
{\sc HII-CHI-mistry} method ({\sc HCm}) of \citet{PM2014}.
Our main conclusions are the following:

\begin{enumerate}
  \item Both methods of abundance determination confirm negative oxygen
  abundance gradients for the majority of galaxies.  Median values of the oxygen
  abundance gradients are $-0.06~\text{dex}/R_\text{e}$ according to the R calibration
  and $-0.03~\text{dex}/R_\text{e}$ for the {\sc HCm} method.
  
  \item The median value of the N/O gradient is also negative,
  $-0.12~\text{dex}/R_\text{e}$ for the R calibration and $-0.18~\text{dex}/R_\text{e}$ for
  the {\sc HCm} method, which is consistent with previous studies
  \citep{Pilyugin2004,PerezMontero2016,Belfiore2017}.

  \item Both methods of the abundance calculation show the correlation between
  the O/H gradient and the stellar mass of a galaxy.  This relation is
  non-linear, with the steepest average gradients around
  $\log(M/M_\sun) \approx 10.0$ and flatter average gradients for galaxies with
  higher and lower masses, as has been previously shown by \citet{Belfiore2017}
  using calibrations of the $R_{23}$ and R3N2 parameters.

  \item The relation between the N/O gradient and the stellar mass is non-linear
  with, on average, the steepest gradients in the intermediate mass range, 
  flatter gradients for galaxies with high masses, and the flattest gradients 
  for low-mass galaxies.  However, the general trend of steepening N/O gradient
  for higher masses remains consistent with the result obtained by
  \citet{Belfiore2017}.
  
  \item Massive galaxies with $\log(M/M_\sun) > 10.25$ show no significant
  correlation between the slopes of either the O/H or N/O gradients and the
  galaxy mean stellar age, as traced by the $D(4000)$ index.
  For galaxies of lower masses, O/H gradients are steeper on average for
  intermediate values of $D(4000)$ and flatter for low and high values of
  $D(4000)$.
  We interpret this behaviour as an evolution of the oxygen abundance gradient
  with the age of the stellar population when young stellar systems have a flat
  oxygen abundance gradient, which becomes steeper with time up to the minimal
  value.
  After this point the oxygen abundance gradient again becomes flatter with time.
  This scenario can be naturally explained by the inside-out growth of galactic
  disks.
  The N/O ratio gradients demonstrate a similar behaviour for the N/O gradients
  derived by both the R calibration and the {\sc HCm} method.

  \item The slopes of the O/H and N/O gradients are on average flatter in
  galaxies with a positive $D(4000)$ gradient, as compared to those with a
  negative $D(4000)$ gradient.

\end{enumerate}

%==========================
\section*{Acknowledgements}

We are grateful to the referee for his/her constructive comments. \\
The work has been performed under the Project HPC-EUROPA3 (INFRAIA-2016-1-730897),
with the support of the EC Research Innovation Action under the H2020 Programme; 
in particular, the authors gratefully acknowledge the support of the
Instituto de Astrof\'{\i}sica de Andaluc\'{\i}a, the computer resources and
technical support provided by the Barcelona Supercomputing Center (BSC). \\
I.A.Z, A.V.S. and M.S. acknowledge support by the National Academy of Sciences of Ukraine under the Research Laboratory Grant for young scientists No. 0120U100148.\\
J.V.M., E.P.M. and S.D.P. acknowledge the support by projects
``Estallidos7'' PID2019-107408GB-C44 (Spanish Ministerio de Ciencia e Innovacion),
the Junta de Andaluc\'{\i}a for grant EXC/2011 FQM-7058 and the Spanish Science
Ministry ``Centro de Excelencia Severo Ochoa Program'' under grant SEV-2017-0709.\\
M.S. was partly supported by the Deutsche Forschungsgemeinschaft
(DFG, German Research Foundation) Project-ID 138713538,
SFB 881 ("The Milky Way System") and by the Volkswagen Foundation under
the Trilateral Partnerships grant No. 97778.\\
M.S. acknowledges support by the Fellowship of the National Academy
of Science of Ukraine for young scientists 2020-2022.\\
S.D.P. is grateful to the Fonds de Recherche du Qu\'{e}bec - Nature et Technologies.\\
Funding for the Sloan Digital Sky Survey IV has been provided by the
Alfred P. Sloan Foundation, the U.S. Department of Energy Office of Science,
and the Participating Institutions.
SDSS-IV acknowledges support and resources from the Center for High-Performance
Computing at the University of Utah.
The SDSS web site is \href{https://www.sdss.org/}{www.sdss.org}. \\
SDSS-IV is managed by the Astrophysical Research Consortium for the 
Participating Institutions of the SDSS Collaboration including the 
Brazilian Participation Group, the Carnegie Institution for Science, 
Carnegie Mellon University, the Chilean Participation Group,
the French Participation Group, Harvard-Smithsonian Center for Astrophysics, 
Instituto de Astrof\'{\i}sica de Canarias, The Johns Hopkins University, 
Kavli Institute for the Physics and Mathematics of the Universe (IPMU) / 
University of Tokyo, Lawrence Berkeley National Laboratory, 
Leibniz Institut f\"ur Astrophysik Potsdam (AIP),  
Max-Planck-Institut f\"ur Astronomie (MPIA Heidelberg), 
Max-Planck-Institut f\"ur Astrophysik (MPA Garching), 
Max-Planck-Institut f\"ur Extraterrestrische Physik (MPE), 
National Astronomical Observatories of China, New Mexico State University, 
New York University, University of Notre Dame, 
Observat\'orio Nacional / MCTI, The Ohio State University, 
Pennsylvania State University, Shanghai Astronomical Observatory, 
United Kingdom Participation Group,
Universidad Nacional Aut\'onoma de M\'exico, University of Arizona, 
University of Colorado Boulder, University of Oxford, University of Portsmouth, 
University of Utah, University of Virginia, University of Washington,
University of Wisconsin, Vanderbilt University, and Yale University.

\bibliography{reference.bib}
 
\end{document}